\documentclass[10pt,letterpaper]{article}
\usepackage[top=0.85in,left=2.75in,footskip=0.75in,marginparwidth=2in]{geometry}

\usepackage[utf8]{inputenc}

\usepackage{cite}
\usepackage{graphicx}
\usepackage{helvet}
\usepackage{hyperref}
\usepackage{amsmath} 
\usepackage{amssymb} 
\usepackage{orcidlink} 
\usepackage{cases}
\usepackage{siunitx}
\usepackage{dcolumn}
\usepackage{bm}
\usepackage{comment}
\usepackage{placeins}

\usepackage{nameref,hyperref}

\usepackage{microtype}
\DisableLigatures[f]{encoding = *, family = * }

\raggedright
\usepackage{changepage}

\usepackage[aboveskip=1pt,labelfont=bf,labelsep=period,singlelinecheck=off]{caption}

\makeatletter
\renewcommand{\@biblabel}[1]{\quad#1.}
\makeatother

\usepackage{lastpage,fancyhdr,graphicx}
\usepackage{epstopdf}
\fancyheadoffset[L]{2.25in}
\fancyfootoffset[L]{2.25in}

\usepackage{color}

\definecolor{Gray}{gray}{.25}

\usepackage{graphicx}

\usepackage{sidecap}

\usepackage{wrapfig}
\usepackage[pscoord]{eso-pic}
\usepackage[fulladjust]{marginnote}
\reversemarginpar

\begin{document}
\vspace*{0.35in}

\begin{flushleft}
{\Large
\textbf\newline{Polarization Plasticity of Ferroelectric Nematics:\\
a case of Electrostatic Frustration in Simple Planar Electrode Cells}
}
\newline
\\
Stefano Marni\textsuperscript{1,X},
Federico Caimi\textsuperscript{1,X,*},
Luca Casiraghi\textsuperscript{1},
Jordan Hobbs\textsuperscript{2},
Calum J. Gibb\textsuperscript{3},
Richard Mandle\textsuperscript{2,3},
Giovanni Nava\textsuperscript{1},
Tommaso Bellini\textsuperscript{1,*}
\\
\bigskip
\bf{1} Affiliation: Medical Biotechnology and Translational Medicine Department, University of Milan, 20133, Milano, Italy
\\
\bf{2} Affiliation: School of Physics and Astronomy, University of Leeds, Leeds, UK, LS2 9JT
\\
\bf{3} School of Chemistry, University of Leeds, Leeds, UK, LS2 9JT
\\
\bigskip
X equally contributed.\\

* federico.caimi@unimi.it, tommaso.bellini@unimi.it

\end{flushleft}

\section*{Abstract}
Because of their spontaneous bulk polarization, ferroelectric nematic liquid crystals can be easily brought by surface coupling or confinement in a state in which the accumulation of bound charges becomes incompatible with polar order, creating frustration. This condition also occurs in the simplest cells with parallel uncoated metal electrodes where we find that the polarization charge accumulating on the electrodes is proportional to the applied voltage for $\Delta V< V_{sat}$, a threshold value independent of cell thickness. We show that below $V_{sat}$, frustration drives the system into a regime of polarization plasticity, in which nematic ordering is preserved and the nematic director remains perpendicular to the electrodes, while instead the polarization is reduced to cancel the internal electric field.  In this regime, the kinetics of bound charge reversal becomes independent from $\Delta V$. The observations are consistent with a model in which the system splits into antipolar tubular domains of mesoscale diameter extending from electrode to electrode, in which the polarization retains its unconstrained equilibrium value, separated by pure polarization-reversal walls.


\section*{Introduction}
Spontaneous polar order is a striking form of collective organization in condensed matter. While ferroic order has been studied for over a century in 3D ferroelectric and ferromagnetic crystals and for decades in 1D crystals (the ferroelectric Smectic C* phase), the recent discovery of the ferroelectric nematic ($N_F$) phase \cite{First_principles_Chen2020, Nishi_DIO,Mandle2017,Lavrentovich2020} has opened the new frontier of a fully liquid material with macroscopic polarization.
Freed from crystal periodicity constraints, the polar ordering in $N_F$ is uncoupled from any specific direction in the 3D space. Thus, unlike all other ferroic materials, $N_F$ cannot self-sustain depolarization field in any significant proportion, a property that makes $N_F$ behave remarkably differently from ordinary nematic Liquid Crystals (LCs). A key manifestation of this difference is the behavior of $N_F$ in contact with solid surfaces. 

The presence of a spontaneous polarization implies that bound charges accumulate on any surface not parallel to $P$. Because polar ordering in the $N_F$ phase involves nearly all molecules~\cite{First_principles_Chen2020}, any component of $\vec{P}$ normal to confining dielectric walls produces a large surface charge density $\sigma_b = \vec{P} \cdot \hat{n}$, which in turn generates strong uncompensated depolarization fields. In solid ferroelectric and ferromagnetic materials, and to some extent also Smectic C*, polarity is preserved even in the presence of adverse fields.
In contrast, because of its fluidity, $N_F$ fluid responds to any internal electric field through local or global reorientation of $\vec{P}$, until the fields are suppressed.

A clear manifestation of this behavior is the cancellation of electric-field components perpendicular to $N_F$-dielectric interfaces. This phenomenon, which we previously termed ``ferroelectric superscreening,'' makes $N_F$ phases extremely sensitive to confinement geometry and enables the guiding of polarization and electric fields along curved micro-channels \cite{Superscreening_Caimi2023}. 
In other dielectric-bounded geometries, such as droplets squeezed between parallel glass plates, interfacial charge accumulation is mitigated by the formation of in-plane striped domains with antiparallel $\Vec{P}$ \cite{First_principles_Chen2020,Soliton_Oleg_Basnet2022,jet_domains_Marni2024}, by twisting along the $z$ axis \cite{domains_Lavrentovich2025}, and by rotations of $\Vec{P}$ near the droplet boundaries \cite{jet_domains_Marni2024} or around air bubble inclusions \cite{First_principles_Chen2020}. As a result, a uniform polar alignment is rarely observed.

Quite different is the behavior of $N_F$ when in contact with the metal surfaces of electrodes. In this case, $\sigma_b$ is mirrored in the electrode by an equivalent amount of free charges $\sigma_e \approx \sigma_b$. The large value of $\sigma_b$ makes the presence of any even tiny separation gap $\delta$ between the planes of bound and free charges a key element affecting voltages and electric fields applied on $N_F$. Indeed, with $\sigma_b \approx$ 5 $\mu$C/cm$^2$ and $\delta \approx$ 1 nm, the voltage drop $V_I$ relative to the insulating layer between electrode and material becomes  $V_I \approx \sigma_b \epsilon_0 /\delta \approx2 V$, a quantity comparable with the minimum applied voltage in many studies of $N_F$
\cite{fibers_Jarosik2024,Osipov_PRE_Nikolova2025,Tunable_Podoliak2025}. Although such experimental situations were investigated on ferroelectric smectics in the 90s \cite{Clark2000_smC_ferro}, early studies on $N_F$ didn't include $V_I$ in the analysis since  in conventional nematics this quantity can be neglected. The relevance of $V_I$ was pointed out \cite{Diel_Clark2024} and included in the most recent dielectric studies \cite{Diel_Adaka2024,Erkoreka_dielec}. Voltage drop at the electrodes has also been recently recognized as the cause of negative capacitance observed in thin $N_F$ sandwich electrode cells \cite{Dhakal2025_neg_capacitance}.

For each electrode, $V_I$ is maximized when the polar axis is normal to the electrode, reaching a limit value of $V_I^{max} = P \epsilon_0 /\delta$. This means that when the applied voltage $\Delta V$ is $\Delta V  = V_{sat} = 2 V_I^{max}$, the entire voltage drops in the vicinity of the electrodes, with $\vec{P}$ and optical axis oriented perpendicular to the electrodes and negligible internal electric field $\vec{E}_{NF}$.  When $\Delta V <  V_{sat}$, a frustrating condition is realized since in this condition maintaining $\vec{P}$ aligned perpendicular to the electrodes would lead to a $\vec{E}_{NF}$  parallel and opposite to $\vec{P}$. The ``polarization external capacitance Goldstone'' (PCG) reorientation mode introduced in Ref. \cite{Diel_Clark2024} solves the problem by assuming that  $\vec{P}$ rotates away from the normal to the electrodes, thus lowering $\sigma_b$ until $\vec{E}_{NF}$ is canceled and thus compatible with the orientation of $\vec{P}$.  While this compensation mechanism can possibly describe the $N_F$ behavior in some thin cells, it cannot be general since rotation implies depositing charges on the lateral dielectric walls of the cell,  generating strong uncompensated fields. At the same time, the splitting of the phase into micro-domains undergoing incoherent rotations around distinct axes would also result in local accumulation of charge and thus large local fields, a condition maybe acceptable as a transient state but not as the equilibrium state when $\Delta V <  V_{sat}$.

In the absence of apparent solutions to these frustrating conditions of conflicting polar order and depolarizing field, and since the condition $\Delta V <  V_{sat}$ is of general relevance in studies and application of $N_F$ phases, we performed electrical and optical (linear and nonlinear) experiments in this regime in planar parallel electrode cells with thicknesses ranging from tenth of microns to millimeters. We find that the $N_F$ phase can enter a regime of ‘Polarization Plasticity’, which we introduce here, in which the effective polarization self-adjusts to cancel the internal electric field while preserving nematic order. The equilibrium and kinetic behavior indicate that the plasticity is achieved through the splitting of the system in tubular domains of opposite polarity connected by pure polarization-reversal walls \cite{Lavrentovich2020}, a configuration that in the language of classical ferroics would correspond to elongated Weiss domains bounded by Ising-like walls and spanning the cell thickness \cite{Lavrentovich2020}. 

\section*{Results}

In this work we explored a variety of cells belonging to two families (see Table \ref{tab:cells}): bulky (B) cells, having sizes is in the $mm$ range in all three dimensions and bare stainless steel electrodes, enabling accurate charge measurements, and optical  (O) thin cells  with in-plane switching type electrodes of different materials for characterization by bright-field and polarized transmission microscopy, Second Harmonic Generation (SHG) and also, in the thickest O-cell, charge measurements. Cells were filled with DIO, synthesized as in Ref. \cite{Hobbs2024} whose $N_F$ phase is in the temperature interval 44 $^\circ C < T < 69  ^\circ C$ \cite{Nishi_DIO}. All experiments have been performed at $T \approx 60^\circ C$. Further details on cells preparation are reported in the Materials and Methods section.

\begin{table}[h!]
\caption{\label{tab:cells} \textbf{Bulky (B) and Optical thin  (O) used in this study} d: electrode distance; h: cell height; w: electrode length; S: electrode surface.}

\begin{tabular}{|l | l | l | l| l | l|} 
 \hline
 \# & d (mm) &h(mm)&w(mm)&S=hw (mm\textsuperscript{2})&Electrodes\\ [1ex] 
 \hline
 B1&0.8 & 2.4 & 2.0 & 4.8 & stainless steel \\ [1ex] 
 \hline
  B2&1.1 & 1.0 & 2.0 & 2.0 & stainless steel\\ [1ex] 
 \hline
  B3&2.2 & 1.4 & 2.0 & 2.8 & stainless steel \\ [1ex] 
 \hline
 B4&4.0 & 2.2 & 2.0 & 4.4 & stainless steel \\ [1ex] 
 \hline
 O1 & 3.0 & 35\textperiodcentered10\textsuperscript{-3} & 10 & 0.35 & stainless steel \\ [1ex] 
 \hline
 O2&3.0 & 20\textperiodcentered10\textsuperscript{-3} &10 & 0.2 & aluminum \\ [1ex] 
 \hline
  O3&3.0 & 11\textperiodcentered10\textsuperscript{-3}&15 & 0.17 & ITO \\ [1ex] 
 \hline
\end{tabular}

\end{table}

\subsection*{Equilibrium Properties: Polarization Charges on the Electrodes}
\begin{figure}[ht!]
\centering
\includegraphics[width=0.95\textwidth]{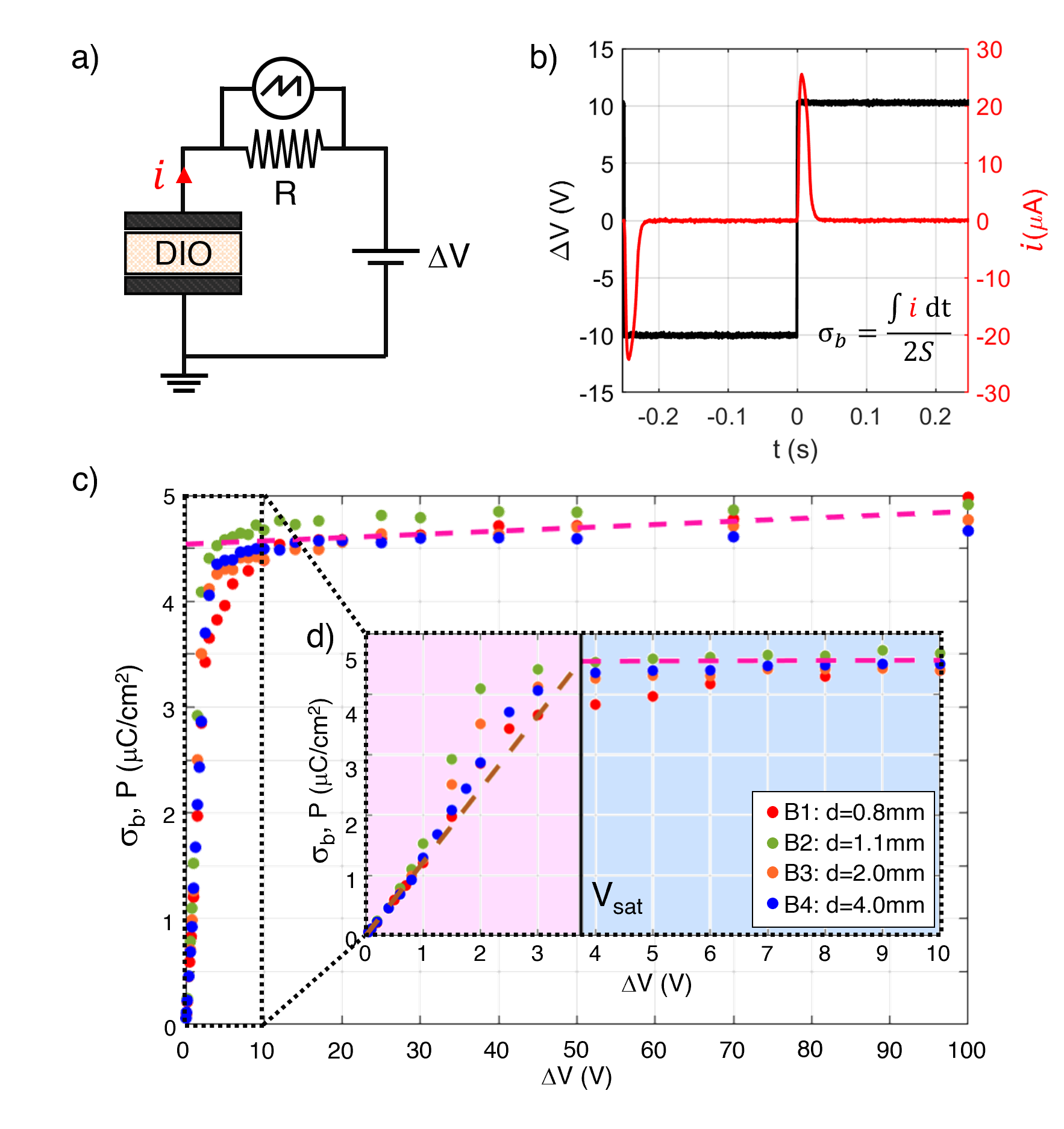}
\caption{\label{fig:fig1_carica} \textbf{Charge measurements} a) Electrical scheme used for the measurement of the current intensity $\mathit{i}$ following voltage reversal. b) Typical $\mathit{i}(t)$ profile (red line, right-hand axis) recorded when applying a squared wave voltage (black curve, left-hand axis). c) Polarization charge density $\sigma_b$ (generally coinciding with $P$, see text) as a function of the applied voltage $\Delta V$ in bulky cells differing in electrode separation $d$. d) Expansion of the low voltage region of panel c. $ V_{sat}$, marking the crossover between Polarization Plasticity (pink shading) and homogeneous $N_F$ polarization (blue shading) is obtained by the intersection of the two linear lines fit at low voltages (0-1V, brown dashed line) and high voltages (10-100V, magenta dashed line). }
\end{figure}

We measured, via polarization-reversal experiments, the polarization charge deposited by the $N_F$ material on the surfaces of the cell electrodes. Measurements are performed by connecting a voltage generator, supplying a square wave of amplitude $\Delta V$, to the cells in series with a probe resistance $R$ (see Fig.~\ref{fig:fig1_carica}a). 
Figure~\ref{fig:fig1_carica}b shows typical $i(t)$ (red lines) flowing after  $\Delta V$ reversal (black lines). The frequency of the square waves was chosen to allow the system to reach equilibrium ($i \approx 0$). Integration of $i(t)$ yields the total variation of surface bound charge $Q = 2 \sigma_e S$ due to polarization reversal (Fig.~\ref{fig:fig1_carica}c), where $S$ is the electrode area.

The surface charge density obtained from the voltage-reversal curves is shown in Fig.~\ref{fig:fig1_carica}c--d as a function of the applied voltage $\Delta V$, for bulky cells with different electrode separations $d$. For $\Delta V > 10\,\mathrm{V}$, we obtain $\sigma_b = 4.6\,\mu\mathrm{C/cm^2}$, consistent with the value of $P$ for DIO reported in the literature~\cite{Nishi_DIO}. We interpret the small drift of $\sigma_b$ with $\Delta V$, captured by the slope of the dashed magenta line, as an experimental artifact.

Upon decreasing $\Delta V$ below $10V$ (Fig. 1d), $\sigma_b$ decreases in all cells, with nearly identical values, toward $\sigma_b= 0$ for $\Delta V =0$. In particular, at low voltages, a clean linear dependence of $\sigma_b$ on $\Delta V$ is observed (brown dashed line), corresponding to a capacitance $C_{eff}/S= 1.2\mu C/(V\cdot cm^2)$ independent from $d$. This finding supports the notion that two insulating layers are present between the polar fluid and the electrodes, as proposed in Ref.  \cite{Diel_Clark2024}. $ V_{sat}=3.7V$ obtained from the intersection of the two linear fits of $\sigma_b$ $vs. \Delta V$ at the large and small $\Delta V$ marks the crossover between two regimes (pink $vs.$ blue shadings). A similar $P(\Delta V)$ behavior was observed by Nikolova et al. \cite{Osipov_PRE_Nikolova2025} using a different cell geometry and material. Their crossover value ($\approx 2.5V$) is compatible with  our measured $V_{sat}$.

\subsection*{Equilibrium Properties: Birefringence and Second Harmonic Generation}

\begin{figure}[ht!]
\centering
\includegraphics[width=0.75\textwidth]{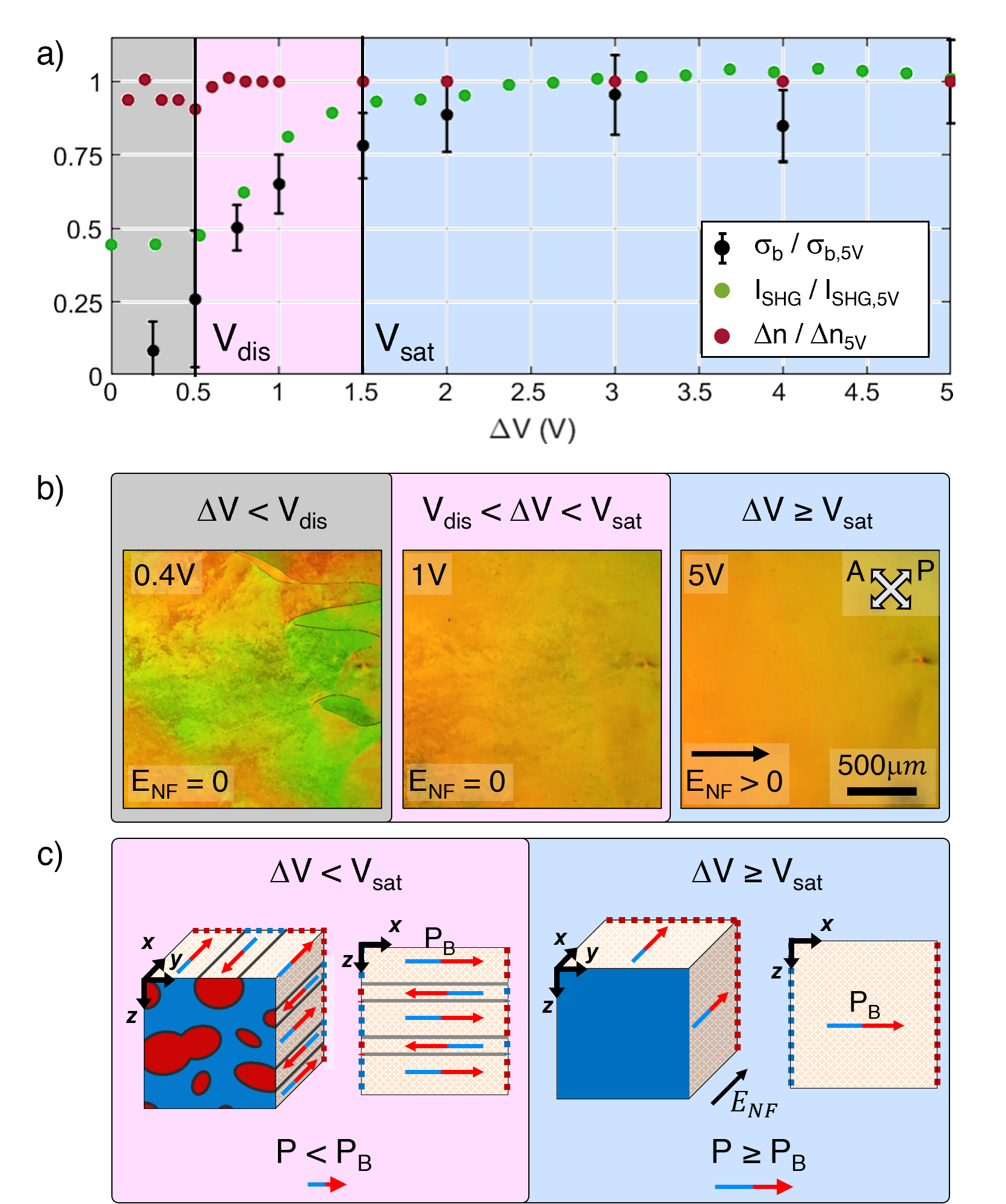}
\caption{\label{fig:fig3_ottica} \textbf{Optical characterization of the O3 cell} a) Comparison of polarization charge density $\sigma_b$ (generally
coinciding with $P$ , see text), SHG intensity $I_{SHG}$, and birefringence $\Delta n$ as a function of $\Delta V$. Each quantity is normalized over their values at $\Delta V=5V$. Color shadings mark the three regimes of voltage described in the text. In pink the polarization plasticity range.  b) PTOM images through cross polarizers P and A (white arrows) at different voltages. Black arrow: direction of the applied electric field E.  c) Axonometric and lateral view of schematic representations of the polarization arrangement in $N_F$ confined between electrodes with $\Delta V <V_{sat}$ and $\Delta V >V_{sat}$(left and right hand side panel, respectively). Red and blue regions indicate surface of positive and negative bound charge accumulation, respectively. In the left hand side panel we sketch the proposed antipolar tubular domain structure that enables polarization plasticity.}
\end{figure}

To better explore the linear decrease of $\sigma_b$ in the low voltage regime, we repeated the charge measurements in the thin cells, to enable the comparison between electrical and optical observations. Although the data are much noisier, the $\sigma_b(\Delta V)$ behavior remains the same as in bulky cells, but with a different $V_{sat}$, indicating a dependence on geometry and electrode material. Data for the O3 cell are shown in Fig. \ref{fig:fig3_ottica}a (black dots). All data sets in the figure are normalized to their value at large $\Delta V$ to allow comparisons. 

Fig. \ref{fig:fig3_ottica}b shows three transmission microscope images of the O3 cell through crossed polarizers. We find that at very low voltages ($\Delta V < V_{dis} \approx 0.5 V$ - gray shaded region) the cell develops irregular textures and color indicating the presence of twisted structures and domain walls. For $\Delta V >  V_{dis}$ the cell takes the appearance of a standard homogeneous nematic with some minor granularity that disappears as $\Delta V$ increases. 

In the same cell we measured the optical birefringence $\Delta n$ with a Berek compensator, which we plot in Fig. \ref{fig:fig3_ottica}a (dark red dots). In line with color observation we find $\Delta n$ to be constant for $\Delta V > V_{dis}$. Measurements at $\Delta V < V_{dis}$ yield slightly reduced and more irregular $\Delta n$ consistent with the loss of homogeneity. In this condition birefringence has been taken as the average of various regions of the cell.

We performed SHG measurements on the same O3 cell using an optical setup built in-house (see Section 1 of Supporting Information and Fig. S1). The resulting data are also shown in Fig. \ref{fig:fig3_ottica}a (green dots). We find the SHG intensity $I_{SHG}(\Delta V)$ to behave similarly to $\sigma_b(\Delta V)$: constant above $V_{sat}$ and dropping to about half its large $\Delta V$ value at $V \approx V_{dis}$. 

The combination of the three measurements shown in Fig. \ref{fig:fig3_ottica}a indicates that the ordering of the $N_F$ phase in planar parallel electrode cells can be split into three regimes: 
\begin{itemize}
    \item a large $\Delta V$ regime ($\Delta V > V_{sat}$, light blue shading), where the polar axis is uniformly oriented perpendicular to the electrodes, so that $\sigma_b = P$. In this regime, (i) $P = P_B$, where $P_B$ is the bulk thermodynamic equilibrium value of $P$, repeatedly measured and reported in literature; (ii) $P$ does not depend on $\Delta V$ (but only on T); (iii) the electric field experienced by the $N_F$ phase is $E_{NF} = \Delta V_{NF}/d =(\Delta V - V_{sat})/d > 0$ and directed as $P$. 
    \item an intermediate regime ($V_{dis} < \Delta V < V_{sat}$, pink shading) in which the nematic order - and thus the polar order - is perpendicular to the electrode, so that $\sigma_b = P$. Here, however, $P$ has a reduced value ($P< P_B$) and it is proportional to the applied voltage ($P \propto \Delta V$) independently from the electrode separation, a finding that demonstrates the cancellation of the internal field ($E_{NF}=0$, see discussion). In this regime the system modifies its effective polarization while preserving the nematic alignment. We define this behavior as ``Polarization Plasticity''.
    \item a low voltage regime ($\Delta V < V_{dis}$, gray shading) in which the director is still largely perpendicular to the electrodes (small decrease in birefringence), but macroscopic distortions are observed.
\end{itemize}

The observation of polarization plasticity in $N_F$ raises the question of the micro or mesoscopic structuring enabling this behavior. One possible scenarios is that of a microscopic reduction of the polar ordering, as if $N_F$ was brought close to a $N_F - N$ phase transition. An alternative scenario is that the system splits into a mosaic of tubular domains with opposite polarity and $P=P_B$, separated by the nematic order-preserving walls, as sketched in Fig. \ref{fig:fig3_ottica}c. In this scenario, the measured $P$ results from the ratio of domains with positive and negative polarity.

\subsection*{Kinetics upon Voltage Reversal: Regimes of Charge Reversal}

\begin{figure}[ht!]
\centering
\includegraphics[width=0.65\textwidth]{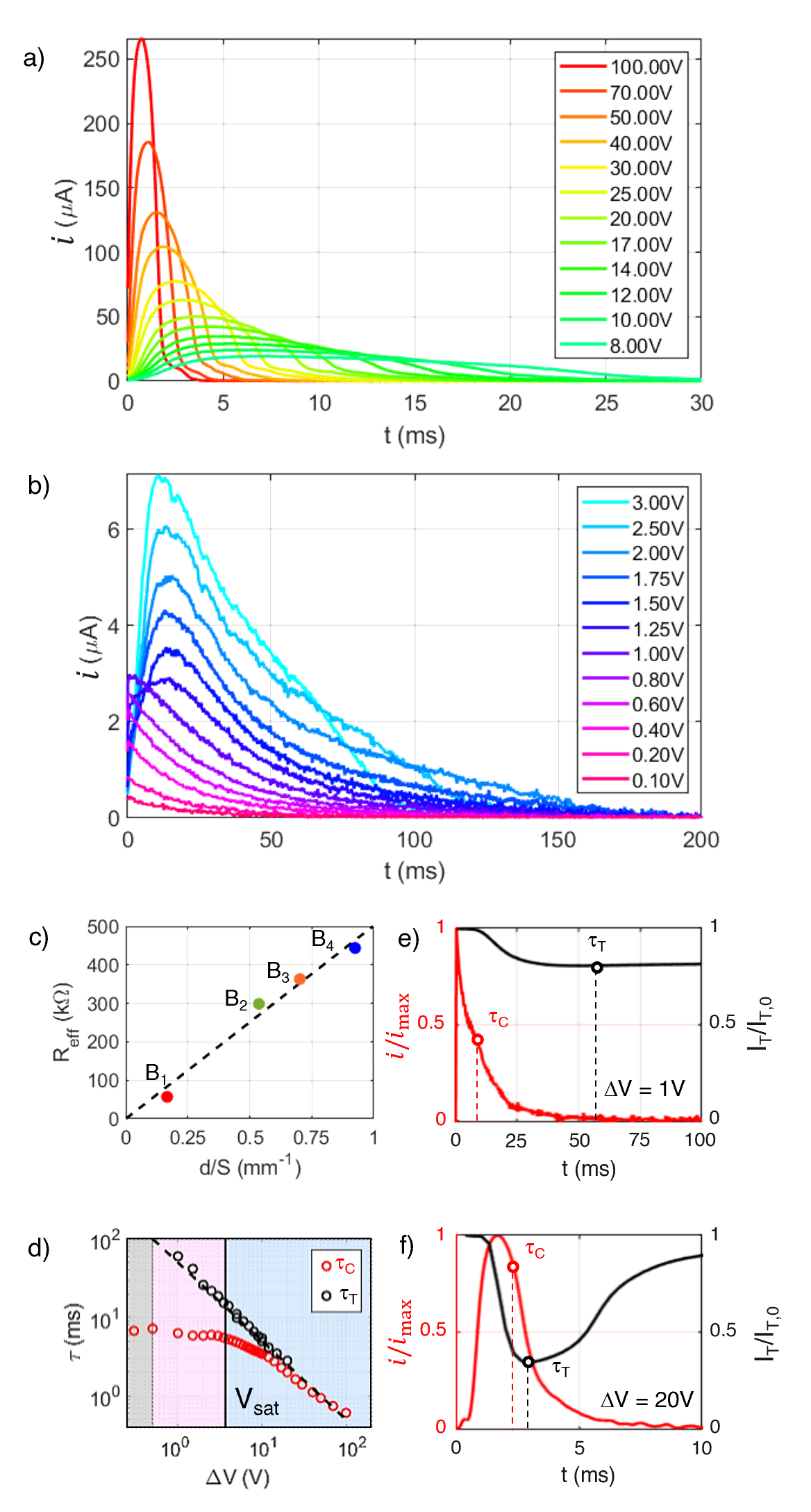}
\caption{\label{fig:fig2_cinetiche}\textbf{Kinetics upon voltage reversal} a-b) $\mathit{i}(t)$ curves measured in the B4 cell with various voltages. c)  Effective resistance $R_{eff}$ determined from the exponential decay time $\tau$ in the bulky cell family, plotted against the shape factor $d/S$. Dashed line: linear fit. d) Discharge time $\tau_c$ (red dots) and optical response time $\tau_T$ (black dots) as a function of $\Delta V$, in the O1 cell. Dashed line: $\Delta V^{-1}$ slope fit to the  $\tau_T$ data. e-f) comparison of the current and optical transmission curves normalized to their largest value as a response to voltage reversal with $\Delta V =$ 1V and $\Delta V =$ 20V, respectively.} 
\end{figure}

Additional information about the polarization plasticity is provided by the time dependence of the charge flow after voltage reversal that also indicate the presence of two regimes. Data from bulky cells show that for $\Delta V > V_{sat} \approx 3.7 V$ (Fig. \ref{fig:fig2_cinetiche}a) the $\mathit{i}(t)$ curves are bell shaped, with peak position and width that decrease with increasing voltage, in line with previous observations \cite{First_principles_Chen2020}.

At lower voltages, ($\Delta V < V_{sat}$, Fig. \ref{fig:fig2_cinetiche}b), the peak position of the $\mathit{i}(t)$ curves is weakly dependent on $\Delta V$, with the whole shape changing with decreasing $\Delta V$ and becoming, for $\Delta V \le 1 V$, well described by exponential decays having a $\Delta V$-independent decay time $\tau$ (see Fig. S2 in Section 2 of Supporting Information). 
In the case of the B4 cell (whose $\mathit{i}(t)$ are shown in the figure), $\tau=24ms$. By interpreting the exponential decays as RC discharge processes, and after having checked that $\tau$ does not depend on the probe $R$ of the circuit, we determined the effective resistance $R_{eff}=\tau/C_{eff}$, based on the effective capacitance determined above. 

Fig. \ref{fig:fig2_cinetiche}c shows the values of $R_{eff}$ for the family of bulky cells plotted as a function of the shape factor $d/S$. Since the dependence is nearly linear, from the slope (dashed line) we can determine the effective resistivity of the polar fluid during the reversal process $\rho_{NF} \approx 500 \Omega \cdot m$. This value is close to the one obtained by dielectric spectroscopy measurements on DIO \cite{Erkoreka_dielec}.

\subsection*{Kinetics upon Voltage Reversal: Comparison between Optical and Charge Switching Times}

$\mathit{i}(t)$ measurements performed on the O1 cell, the thickest of the optical cells, confirm the same behavior observed in the B cells, although curves are noisier because of the much smaller S (see Table I). From $\mathit{i}(t)$  we extracted the discharging time $\tau_c$, defined as the time after voltage reversal at which half of the flowing charge $Q$ has flown in the circuit, which we plot  in Fig. \ref{fig:fig2_cinetiche}d (red dots) as a function of $\Delta V$. In line with what observed above for the bulky cells, when $\Delta V \lesssim V_{sat}$, $\tau_c$ becomes weakly $\Delta V$ dependent and saturates to a constant value for $\Delta V \rightarrow 0$.

Since the O1 cell enables optical observations, we could
study the time dependence of the transmitted intensity $I_T(t)$ as measured in bright field transmission microscopy (no polarizers) after voltage reversal, as in Ref.  \cite{Superscreening_Caimi2023}. During such process $I_T(t)$ starts from an initial value $I_{T,0}$, it decreases to a minimum and it eventually grows back to $I_{T,0}$. The reduction of $I_T$ reflects the disruption of the uniform nematic alignment into domains in which the polarization and the optical axis incoherently rotate toward the reversed direction. Such incoherent rotation gives rise to an optical inhomogeneous, and thus turbid, state which becomes again transparent when the rotation is completed and uniform alignment restored. From $I_T(t)$  we thus extracted $\tau_T$, the delay after voltage reversal at which the largest turbidity - lowest $I_T$ - is found. More details on $\tau_T$ calculation can be found in Section 3 of Supporting Information.

Fig. \ref{fig:fig2_cinetiche}e and \ref{fig:fig2_cinetiche}f compare $i(t)/i_{max}$ with $I_T(t)/I_{T,0}$ for respectively two choices of $\Delta V$. For $\Delta V > V_{sat}$ (panel f),  both curves are bell-shaped with comparable characteristic times, as also found in \cite{First_principles_Chen2020}, while for $\Delta V < V_{sat}$ (panel e), $\tau_c$  is much shorter than $\tau_T$. By comparing the two panels, it can also be noticed that the minimum of $I_T(t)/I_{T,0}$ is much more pronounced at large $\Delta V$, indicating a larger turbidity. 

A direct comparison between $\tau_T$ and $\tau_c$ measured in the O1 cell as a function of $\Delta V$ is shown in Fig. \ref{fig:fig2_cinetiche}d. For large voltages the two curves collapse, both exhibiting a $\Delta V^{-1}$ trend (dashed black line). Conversely, by lowering $\Delta V$ below $ V_{sat}$, the two trends differ, with $\tau_T$ maintaining the $\Delta V^{-1}$ dependence even at low $\Delta V$. It is remarkable that in this regime we can find $\tau_c\approx 0.1 \tau_T$, indicating that the fluid delivers a significant amount of charge much faster than the time it takes to resolve the optical disordering following voltage reversal. 

The availability of a mechanism to transport bound charges between electrodes that does not generate optical turbidity becomes apparent when exploring sudden variations of $\Delta V$ between values of equal sign. When switching from $\Delta V_1$ to  $\Delta V_2$, both larger than $V_{sat}$, $\sigma_b$ stays constant and no optical event is observed. Conversely, if $\Delta V_1<V_{sat}$ and $\Delta V_2>V_{sat}$, the transition from $\Delta V_1$ to $\Delta V_2$ involves bound charge displacement (from $\sigma_b = C_{eff} \Delta V_1 /S$ to $\sigma_b = P$) but no observable turbidity. If instead the transition is from  $\Delta V_2$ to $\Delta V_1$, a global perturbation generating optical turbidity is observed, with features similar to those observed upon voltage reversal (see Fig. S4, in Section 4 of Supporting Information, for the case $\Delta V_1=0.5V$, $\Delta V_2=5V$).

\section*{Discussion}

\subsection*{Equilibrium}

\begin{figure}[ht!]
\centering
\includegraphics[width=0.85\textwidth]{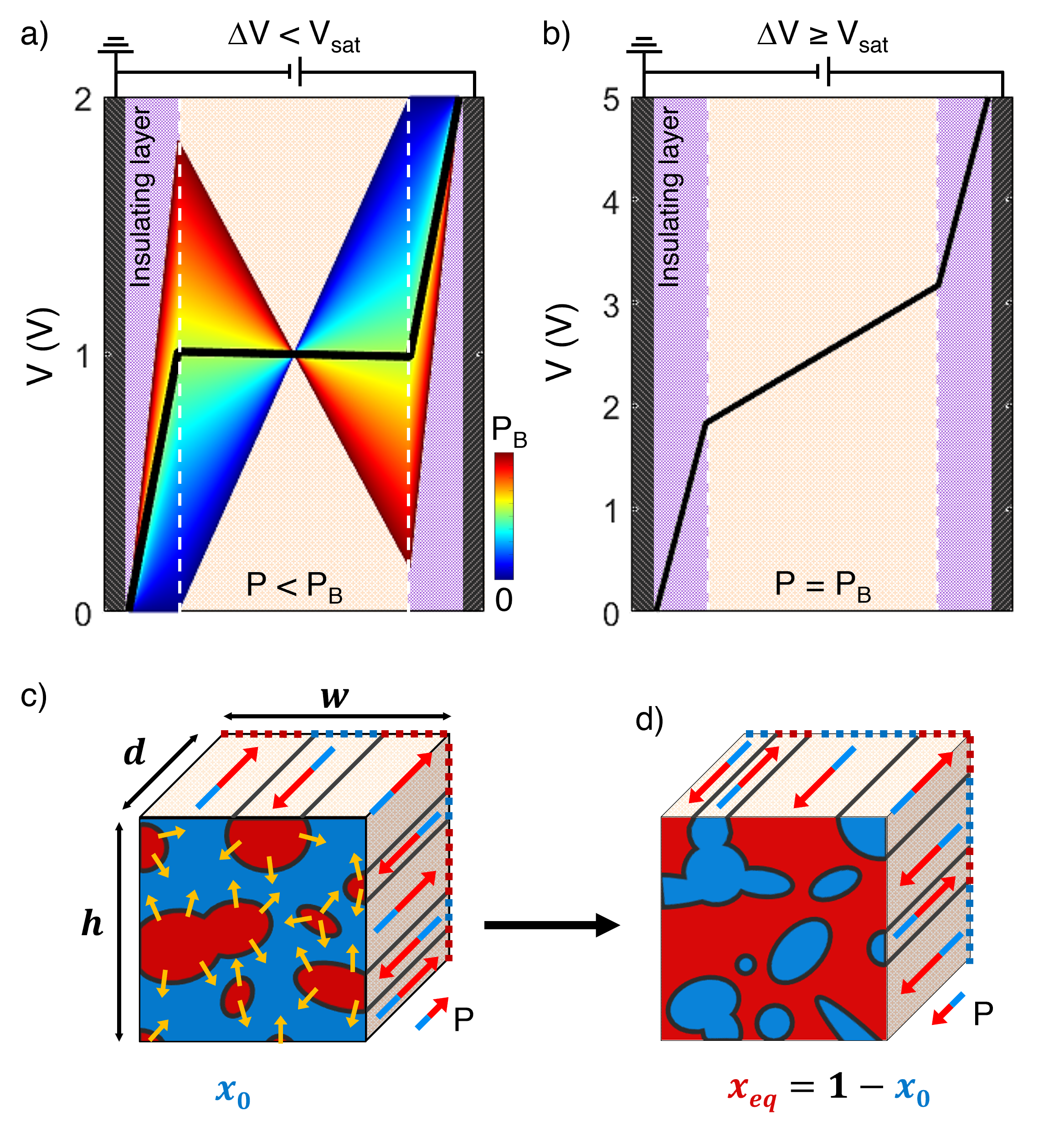}
\caption{\label{fig:fig4_modello} \textbf{Electrostatics and structure in the polarization plasticity state} a-b) Thick black line: voltage at equilibrium as a function of the distance from the grounded electrode for $\Delta V< V_{sat}$ (a) $\Delta V \geq V_{sat}$ (b). Colors in (a) indicate the voltage profile across the cell that would be present if $P$ was larger (yellow to red) or smaller (green to blue) than its equilibrium value. c-d) same sketch of the Antipolar Tubular Domain structure as in Fig. 2c. Yellow arrows indicate the motion of the domain walls by which the minority fraction  (red) expands to reverse the sign of $P$, as in the transition following voltage reversal.}
\end{figure}

We have found that for $\Delta V < V_{sat}$ all $\sigma_b$ $vs.$ $\Delta V$ data obtained in bulky cells with a wide range of electrode separation ($0.8mm-4mm$) collapse on a common $\sigma_b=\Delta V C_{eff}/S$ line (Fig. \ref{fig:fig1_carica}c). This finding demonstrates that  voltage drop and electric field within the $N_F$ fluid 
are negligible. In this condition the whole $\Delta V$ drop takes place at the electrodes, as previously suggested  \cite{Clark2000_smC_ferro} (thick black line in Fig. \ref{fig:fig4_modello}a). 

The reduction of $P$ with respect to the bulk value $P_B$ observed for $\Delta V < V_{sat}$ can be understood as the only available solution to avoid instability. Indeed, if $P$ were to maintain its bulk value, the voltage drop across the $N_F$ fluid would be $\Delta V - V_{sat} < 0$, and thus a field opposite to \textbf{$P$} would arise (yellow-to-red values in \ref{fig:fig4_modello} a), leading to an unstable condition.
Symmetrically, if a smaller $P$ was present, a positive field  (green-to-blue values) would promote its growth until the field is canceled. The condition of polarization plasticity can only be found when no electric field is present in the $N_F$.

The value $C_{eff}/S= 1.2\mu C/(V\cdot cm^2)$ extracted from bulky cell data refers to the two electrode capacitors in series, each of capacitance $C_I$. By treating these as planar capacitors, consisting of a dielectric layer of thickness $\delta_I$ and permittivity $\varepsilon_I$ interposed between the $N_F$ and the electrode, we obtain $C_I=2C_{eff}=\varepsilon_I S / \delta_I$.
If we take $\varepsilon_I\approx \varepsilon_0 $ we find $\delta_{I}= 3.6$\AA, indicating that our observations are consistent with a very thin $N_F$-electrode separation (the molecule length is $\ell_{DIO} \approx 20$ \AA), sufficiently thin to be compatible with the intrinsic disorder and/or staggering present in the fluid near the solid surface. Sensitivity to such a small interfacial gap reflects the distinctive properties of the $N_F$: a large bound-charge accumulation $\sigma_b$ combined with fluidity and thus vanishing coercive field. 

The value of $V_{sat}$ depends on the electrode material: from the $P(\Delta V)$ curves in O2 and O3 cells, built with aluminum and ITO electrodes we obtain $V_{sat}= 10 V$ and $V_{sat}= 1.5 V$, respectively. The larger $V_{sat}$ in O2 indicates a thicker insulating layer that we understand as a Al\textsubscript{2}O\textsubscript{3} oxide layer whose typical thickness is of the order of 1-10 nm \cite{Al_oxide}. By considering an electrode gap formed by a  $0.36nm$-thick layer of vacuum (assumed to be general for $N_F$-solid surface interfaces) and by a layer of $Al_2O_3$ ($\varepsilon_{Al_2O_3}=10\varepsilon_0$), we obtain a thickness $\delta_{Al_2O_3}= 6nm$. Interpreting the ITO cell is instead made difficult by the different geometry, with flat surface-bound electrode. 

When $\Delta V\geq V_{sat}$, $P=P_B$ and the voltage drop in $N_F$ is $\Delta V_{N_F}=\Delta V-V_{sat}$ (thick line, Fig. \ref{fig:fig4_modello}b). 
Thus, 
\begin{subnumcases}{P(\Delta V) = }
   P_{B} & $\Delta V\geq  V_{sat}$ \label{eq_P_sotto_DV_sat}
   \\
    \frac{\Delta V}{V_{sat}}P_{B}=\frac{\varepsilon_I}{2\delta_I}\Delta V & $\Delta V\leq V_{sat}$. \label{eq_P_sopra_DV_sat}
\end{subnumcases}

The observation that a condition of reduced $\sigma_b$ is not obtained through a rotation of the polar order - which would lead a charge accumulation on the side walls generating an opposing field - but via a polarization reduction of a well aligned nematic with $P < P_B$ (Eq. 1b) is the core finding of this work. This condition of polar plasticity can only be obtained by inverting a fraction of the molecules while maintaining their nematic order. An unavoidable question is on which length scales such inverted molecules are polarly correlated. One possibility is that the correlations are very short, with molecular lengths, which would correspond to a uniform loss of polar order parameter as if approaching a transition toward a conventional nematic. This scenario appears however energetically unfavorable since uniform nematization would require to overcome the thermodynamic free energy associated to polar order, an estimate of which is the latent heat of the polar-apolar transition, $\sim 100 J/kg$ \cite{Thoen2024}. This value should be compared with the electrostatic energy density  $P_B\cdot V_{sat}/d \sim 0.1 J/kg$, a much smaller quantity. Furthermore, a transition toward a conventional (non-polar) nematic typically brings a significant drop in birefringence \cite{Brown2021}, which we do not observe.

A more realistic alternative is that the polar correlation extend on the mesoscale, giving rise to a splitting of the system into tube-shaped domains having local polarization $P_B$ reversed polarity. We sketched such ``Antipolar Tubular Domains'' (ATD) scenario in Fig. 2c. With respect to the uniform reduction of $P$, in which the largest majority of the molecules would be involved in antipolar interactions, the ATD structure has the advantage of  reducing the antipolar interface and thus the associated energy cost. Also, because of the electrode-to-electrode tubular structure, the charges accumulating an the bases are compensated by the free positioning of the charges on the electrodes. 
 
While the combination of reduced P and unmodified optical axis and birefringence demonstrates polarization plasticity, no optical feature indicating inter-domain walls is observed. We interpret this feature as an evidence that the interface between ATD are Pure Polarization Reversal walls observed in $N_F$ between domains of opposite polarity  \cite{Lavrentovich2020}, and in other polar phases such as in $SmA_F$ \cite{Chen2022}, in splay nematic antiferroelectric \cite{Sebastian2022}  and in $Sm Z_A$ phases \cite{Chen2023}. Indeed, in the need to escape frustration, the system might drift toward anitferroelectric-like structures, which are closer to the $N_F$ phase than the apolar nematic phase \cite{Chen2023}.

\subsection*{Polarity Reversal}

In the PCG relaxation mode approach, the polarity reversal takes place through a coherent rotation of the polarization, resulting in symmetric bell shaped $i(t)$ with peak time $t_p\sim P_{b}\rho_{N_F}d/\Delta V$, as shown in Section 5\textsubscript{b} of Supporting Information . This result, obtained for $\Delta V\gg V_{sat}$, is in semi-quantitative agreement with the observed $i(t)$ in the large voltage regime (Fig. \ref{fig:fig2_cinetiche}a), where inversion takes place through splitting in mesoscopic domains undergoing rotations and where $\tau \propto \Delta V^{-1}$ (Fig. \ref{fig:fig2_cinetiche}d). 

When $\Delta V < V_{sat}$, a second distinct polarization reversal mode becomes active, as indicated by: (i) the change in the shape of the current curves that cross-over from a bell-shaped to a decaying exponential response (Fig. 3b); (ii) the separation between $\tau_c$ and $\tau_T$, with the former deviating from the $\Delta V^{-1}$ scaling (Fig. 3d);  (iii) the decreased turbidity with a very slow recovery at small $\Delta V$ (Fig. 3e and Fig. S4). Being this second reversal mode appearing together with the polar plasticity of $N_F$, it is reasonable to assume that these phenomena are intimately connected. 

The Antipolar Tubular Domains structure described above enables an alternative polarization reversal mechanism. In the ATD structure, the average polarization $P$ is determined by the fraction $x$ of the system with ``positive'' (arbitrarily chosen) polarity: $P(x)=+xP_{B}+(1-x)(-P_{B})=(2x-1)P_{B}$. At equilibrium, $x_{eq}= 1/2 + \Delta V/2 V_{sat}$.  Thus, in the polarization plasticity regime, changes in $P$ can occur via the motion of the domain walls.

Fig. 4c and d sketch the process of polarization inversion via domain wall displacement. After voltage reversal, the previous equilibrium state becomes unstable and the system undergoes a transformation $x_{eq} \rightarrow 1-x_{eq}$. 

As the walls move, a change $\delta x$ in the fractional polarity is produced with a corresponding variation of electrostatic energy $\delta U=-2P_B E_{N_F} v\delta x$, where $v=hwd$ is the cell volume (see Fig. \ref{fig:fig4_modello} c). Thus, the force per unit length acting to bring the walls to the equilibrium position is:
\begin{equation}
\label{eq:f(x)}
    F(x)=-\frac{\delta U}{S\delta x}=-4P_{B} V_{sat}(x-x_{eq}),
\end{equation}
where $S$ is the electrode surface ($S=hw$). The wall motion will be determined by the balance of such force with the contrasting viscosity: $F(x)=- \eta \dot x$, where $\eta$ has the dimension of a surface viscosity.

The very same process determines the current of polarization charges:
\begin{equation}
    i=2P_{b}S \dot x=\frac{\Delta V_{N_F}}{R_{eff}},
\end{equation}
where $R_{eff}$ the effective resistance defined above
Eq.3 maps into the force balance equation provided that $\eta=4P_B^2d\rho _{N_F}$. The quantity $P^2 \rho _{N_F}$ is a rotational viscosity \cite{Diel_Clark2024}. 

By choosing as initial condition $x_0=1-x_{eq}$, as in a voltage reversal experiment, we obtain:
\begin{equation} 
\label{eq_wall_xt}
    x(t)=\frac{1}{2}+\frac{\Delta V}{2 V_{sat}}-\frac{\Delta V}{ V_{sat}}exp\left(-\frac{t}{R_{eff}C_{eff}}\right).
\end{equation}

Thus, the wall dynamics leads to exponentially decaying currents with characteristic time $R_{eff}C_{eff}$, which agrees with the basic features of the observed additional route for polarization inversion,  exponential in shape (Fig. \ref{fig:fig2_cinetiche}b and Fig. S2 in Section 2 of Supporting Information) and with a $\Delta V$-independent characteristic time (Fig. \ref{fig:fig2_cinetiche}d). 

Thus, while at $\Delta V >  V_{sat}$ the polarization inverts through optically detectable rotations taking place in a $t_p\sim \Delta V^{-1}$ time common to optical and electric inversion signatures, at $\Delta V<  V_{sat}$ the situation is richer. According to our proposed ATD model, the system becomes divided in domains of opposite polarities, a situation that makes it possible for both mechanism of polarity inversion to be active: the optically undetectable wall displacement yielding exponential decaying current, and the optically detectable rotation of portions of the domains that are directed opposite to the field. Indeed, the presence of a $\sim \Delta V^{-1}$ decaying mode is by itself an indication that there are portions of the sample where the local polarization is $P_B$ and opposite to the field, so that their inversion time by rotation scales with continuity to the $\Delta V >  V_{sat}$ behavior. 

A strong confirmation of the polarization changes via domain expansion/contraction is given by the response to asymmetric switches described above (Fig. S3, in Section 3 of Supporting Information): $P$ variations via wall displacement is possible when the starting conditions of the system is to be split in ATD (as in $\Delta V_1=0.5V$ to $\Delta V_2=5V$ switches) but not when the system is initially uniformly polarized with $P=P_B$ (as in $\Delta V_2=5V$ to $\Delta V_1=0.5V$ switches). In this latter condition the system thus responds by breaking in domains that undergo rotation.

\section*{Conclusion}

In this work we have explored the behavior of a ferroelectric nematic fluid in a seemingly straightforward condition: between parallel planar electrodes at low voltage.  By comparing the behavior of bulky cells with various electrode distance we have shown that there is a threshold voltage $V_{sat}$ (already reported in previous works \cite{Diel_Clark2024,Osipov_PRE_Nikolova2025}) at which the $N_F$ polarization is at its bulk value $P_B$ and the applied voltage entirely falls at the electrodes because of their intrinsic capacitance. When the voltage applied to the cells is below that threshold, the system becomes frustrated, unable to maintain its bulk polarization while avoiding depolarization fields. We find that in this condition, the $N_F$ phase enters a different state, as testified by equilibrium and dynamic, electric and optical observations. In such a state, the ferroelectric fluid maintains the nematic ordering with unmodified birefringence and no sign of disordering into domains (except at very low voltages) while effectively lowering its average bulk polarization. We have found that in such state, that we have dubbed of ``Polarization Plasticity'', the system modifies not only it polarization but also it kinetic response to electric field variation, with the appearance of a route for polarization inversion faster than $\sim \Delta V^{-1}$. 

We argue that the structure most compatible with the data is that of a system divided into tubular domains of opposite polarity, which we define as ``Antipolar Tubular Domains'' model, separated by pure polarization reversal walls. Indeed, such a structure combines: (i) the tendency of $N_F$ materials to develop, in contiguous phases walls separating domains of opposite polarities as in the Smectic A\textsubscript{F} or splay-nematic phase \cite{Chen2022, Sebastian2022, Chen2023, Rupnik2025}; (ii) the local polarizability equal to the bulk $P_B$, a condition that minimizes energy costs and is coherent with the observation of a remnant $\sim \Delta V^{-1}$ decay time; (iii) the availability of an additional degree of freedom - the motion of domain wall - that gives rise to an alternative switching process that has no optical signature.

The polarization plasticity state is peculiar to ferroelectric nematics, with no counterpart in ferroic solid materials in which the coupling of $P$ to the crystallographic direction enables the formation of history-dependent ordering, such as the one intentionally produced by poling. Polarization plasticity, here demonstrated for a simple configuration, could actually be the general strategy for $N_F$ phases to avoid electrostatic instability in a much wider set of frustrated conditions, including defect, domain walls and splay-inducing boundary conditions. This state, and the associated exponential relaxation upon voltage reversal, could also come into play in the design of applications of $N_F$ for fast-charging processes in energy storage.

\subsection*{Cell preparation}

Bulky cells are constructed using two stainless steel plates as electrodes,  separated by a silicone sheet acting both as a spacer and gasket. The sheet contains a precut mask defining the measurement volume, with the top side open for material insertion. DIO is loaded in the solid phase, heated in an oven at $T=140^oC$ for the minimum time required to melt into the $N$ phase, and quickly cooled in the $N_F$ at $T=60^oC$, where experiments are conducted. The distance between the electrodes $d$, as well as the height $h$ and the width $w$ of the mask are measured with an electronic caliber. The actual filling of the cell has been verified by measuring the DIO mass $m$ after the experiment and computing the volume $v=m/\rho_{DIO}$, where $\rho_{DIO}=1.3 g/cm^3$ \cite{Parton_DIO_density}.

IPS optical thin cells with metallic electrodes are produced using two stainless steel or aluminum foil strips as both electrodes and spacers between two optical glasses. Instead, cells with ITO electrodes are produced by sandwiching two Mylar film spacers between an optical glass slab and an ITO-coated glass slide, from which the electrodes and the optical region are etched. The etching is obtained by masking the electrode regions with silicone pads and depositing a small volume of a 50\% v/v solution of H\textsubscript{2}SO\textsubscript{4} in H\textsubscript{2}O on the exposed area. The slide is then heated on a hot stage at $T=80^oC$ for 5 minutes, carefully washed with H\textsubscript{2}O and air dried. For every thin cell, glass surfaces are treated with hexadecytrimethoxysilane (HDTMS) and left unrubbed to favor a degenerate planar alignment \cite{Caimi21_surfaces}. Silanization is performed by stirring a 2\% v/v solution of HDTMS in a 95:5 ethanol-water mixture, acidified to pH 5 with acetic acid, for 5 minutes at room temperature. The solution is then pipetted onto the slides and, after solvent evaporation, the slides are heated on an hot plate at $T=110^oC$ for 10 minutes and stored at room temperature over night. The slides are eventually rinsed with isopropanol to remove the excess of silane. Cells are assembled with metallic clips and the thickness is measured via Fabry-Perot interferometry technique. DIO is finally loaded by capillarity in the $N$ phase at $T=140^oC$ and immediately cooled in the $N_F$ phase at $T=60^oC$.

\subsection*{Acknowledgments}

The authors would like to thank Prof. Noel Clark for precious discussions. S. M. and F. C. acknowledge support by Fondazione Cariplo, grant Young Researcher no. 2023-1095. R.J.M thanks UKRI for funding via a Future Leaders Fellowship, grant number MR/W006391/1, and the University of Leeds for funding via a University Academic Fellowship. R.J.M. gratefully acknowledges support from Merck KGaA.


\section*{Supporting Information}

The present Supporting Information is composed as follows:
\begin {itemize}
 \item Section 1: SHG measurement setup and Figure S1;
 \item Section 2: RC relaxation curves at low voltages and Figure S2;
\item Section 3 $\tau_T$ extraction from the transmitted intensity curves and Figure S4;
\item Section 4: Response to asymmetric voltage switches and Figure S3;
\item Section 5: Derivation of the kinetic behavior from the PCG model reorientation mode.
\end{itemize} 
\setcounter{figure}{0}
\renewcommand{\thefigure}{S\arabic{figure}}

\subsection*{SI Section 1: SHG measurement}

SHG measurements were performed with a custom-made optical setup (Fig. S1), consisting of a Nd:YAG pulsed laser (pulse duration 10ns, repetition rate 10Hz) at $\lambda=1064nm$ impinging perpendicularly on the optical cells polarized along the $N_F$ director by a Glan-Laser polarizer. The second harmonic signal is collected through a second polarizer parallel to the first and analyzed  with a Ocean USB2000+ spectrometer to selectively measure the intensity of the $\lambda=532nm$ peak, rejecting stray light and possibly minimizing power leakages from the carrier wavelength.  The integrated  SHG peak was measured at various $\Delta V$ to obtain $I_{SHG}(\Delta V)$. For every voltage, measurements were repeated in over 20 zones around the sample and then averaged.

The detected $I_{SHG}$ signal is, by definition \cite{Svelto2010}, the result of the chosen experimental conditions: (i) is a tensorial quantity as it depends on the polarization of the input photons, which summed up create the doubled frequency, and of course from the final polarization analyzed; (ii) it changes periodically with the optical path length as it suffers from the phase matching of the second harmonic signal and the carrier wavelength. To avoid introducing artifacts due to (i) and (ii) we performed SHG measurements on the same thin optical cells we used for charge/optical measurements (the thickness was kept constant) and we employed glan-laser polarizers to carefully control both the input and the output optical polarization to be parallel to the liquid crystal director of $N_F$ 

\begin{figure}[!ht]
\centering
\includegraphics[width=0.85\textwidth]{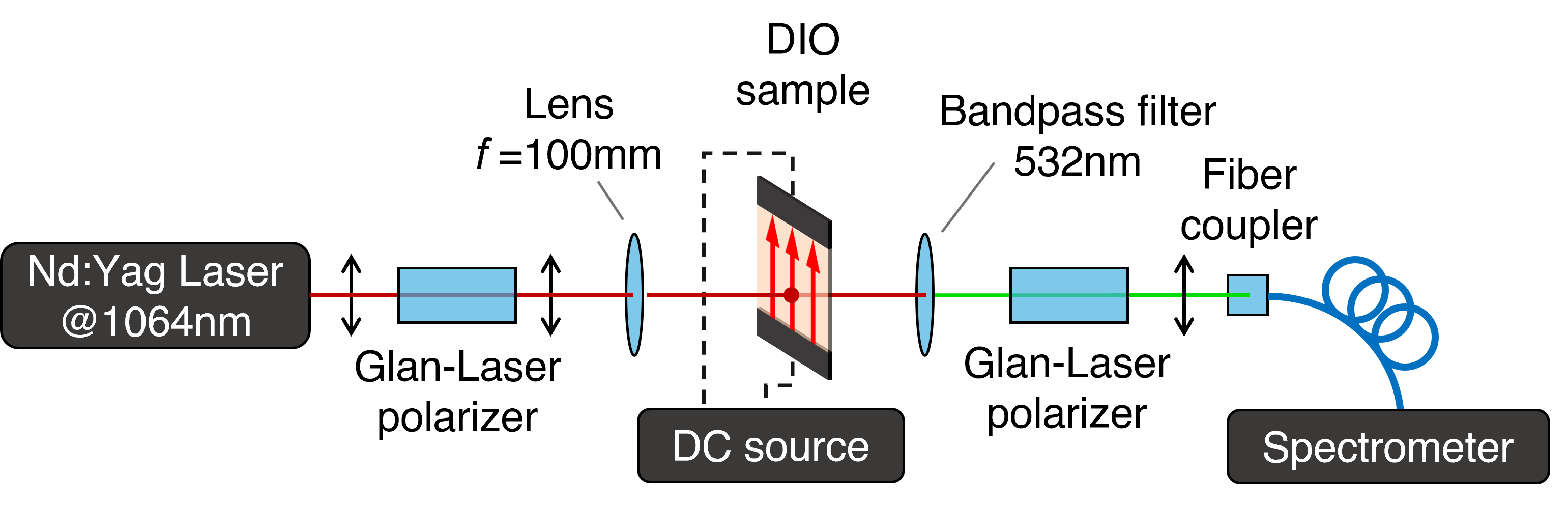}
\caption{\label{fig_apx:shg_setup} \textbf{SHG measurement setup} Schematic representation of the custom SHG intensity measurements setup used in this work.} 
\end{figure}

\newpage
\subsection*{SI Section 2: RC relaxation curves at low voltages}
\label{app_RC}

\begin{figure}[!ht]
\centering
\includegraphics[width=0.75\textwidth]{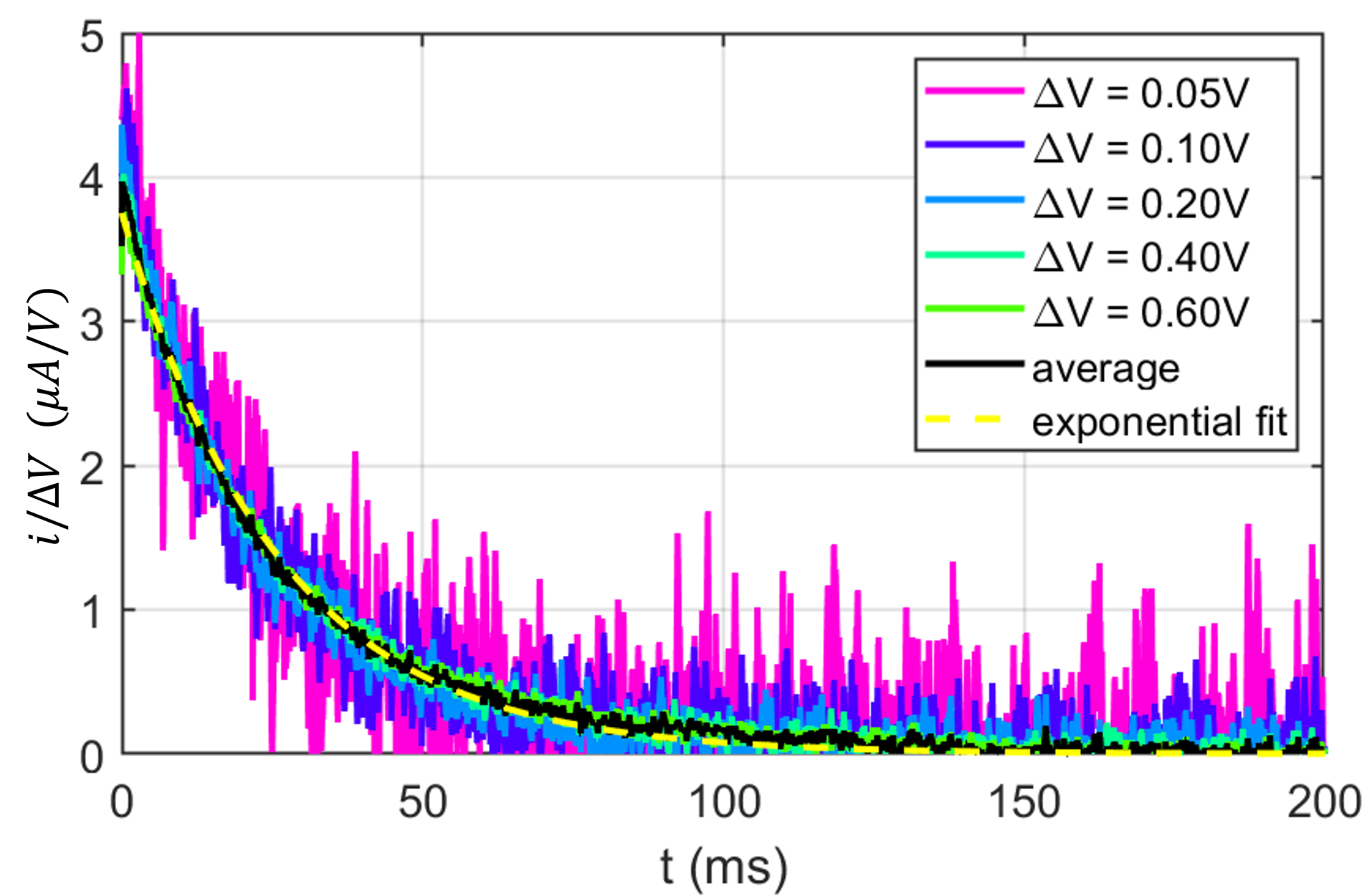}
\caption{\label{fig_app_expo_fit} Current intensity $i(t)$ at low voltages (colored solid curves), normalized on $\Delta V$ in the B4 cell. In black the weighted average of the curves, fitted with an exponential decay (dashed yellow line).} 
\end{figure}

At low voltages the $i(t)$ curves exhibit an exponential decay typical of RC relaxation processes. In Fig. \ref{fig_app_expo_fit}, we show the $i(t)/\Delta V$ curves in the B4 cell at low voltages (colored lines). From such curves, we obtain an average curve (black line), that is well fitted with an exponential decay (yellow dashed curve).

\FloatBarrier
\newpage

\subsection*{SI Section 3: Response to asymmetric voltage switches}
\label{app_switch}
\begin{figure}[h!]
\centering
\includegraphics[width=0.75\textwidth]{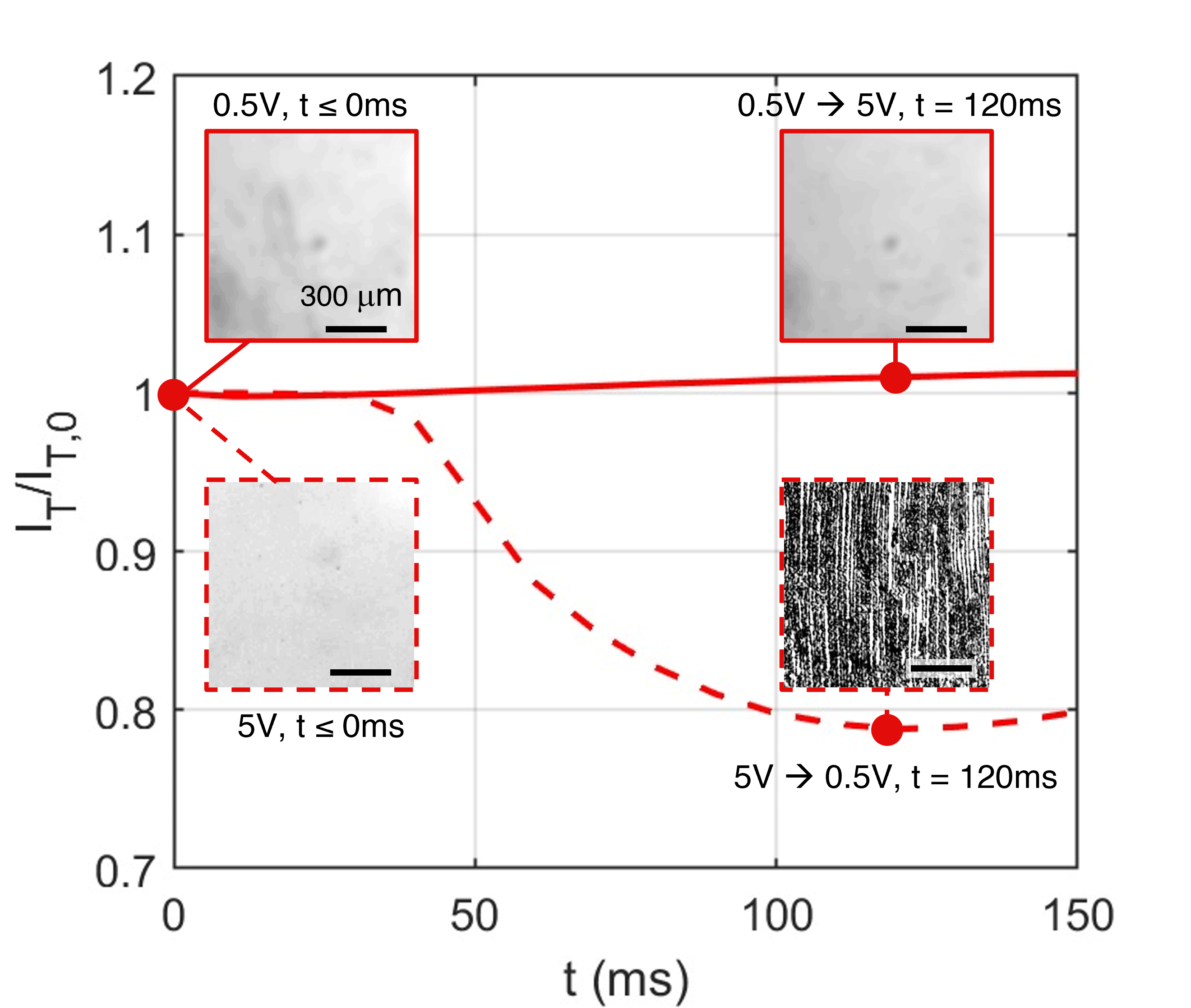}
\caption{\label{fig_apx:app_switch} Transmitted intensity $I_T$ measurements after voltage change across $V_{sat}$, in the O3 cell. Solid line: $I_T$ after the voltage change from $0.5V$ to $5V$; dashed line: $I_T$ after the voltage change from $5V$ to $0.5V$. Pictures represents frame of the two measurements, in equilibrium before the switch and at $t=120ms$, wherein $I_T$ reaches its minimum value in the $5V$ to $0.5V$ measurement.} 
\end{figure}

To better understand the behavior of the polar fluid during domain expansion/contraction we studied the response to asymmetric switches: the most emblematic case of voltage switches from $0.5V$ to $5V$ and vice-versa is reported in Fig. \ref{fig_apx:app_switch}, with the transmitted intensity curves and snapshots of the processes. From $0.5V$ to $5V$ there is not significant perturbation of the fluid, whereas from $5V$ to $0.5V$ the system responds by breaking in domains that undergo rotation.

\FloatBarrier
\newpage

\subsection*{SI Section 4: \texorpdfstring{$\tau_T$}{tau\_T} extraction}

The response to voltage reversal results in a reconfiguration of the polar alignment: the homogeneous $N_F$ medium becomes turbid decreasing the bright field transmitted intensity $I_T$ and then recovers the polar order, coming back to the initial transmitted intensity $I_{T,0}$. We measured $I_T$ with a Teledyne Prime BSI Express camera at 3030 fps, which has been synchronized with a National Instrument PCIe-6323 multifunctional I/O card as voltage source to properly identify and image the voltage reversal moment. After computing $I_T$ as the average signal in the ROI, we obtain maximum turbidity time $\tau_T$ as the moment after the voltage reversal in which $I_T$ reaches its minimum value. In Fig. S4 the normalized bright field transmission intensity curves $I_T(t)$ after the voltage reversal are shown, together with $\tau_T$.

\begin{figure}[h!]
\centering
\includegraphics[width=0.75\textwidth]{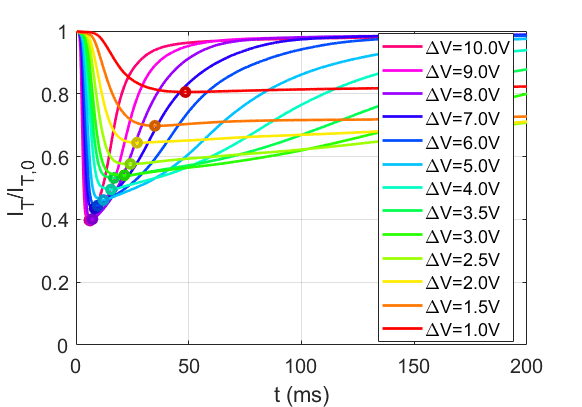}
\caption{\label{fig_apx:turb_vs_time} \textbf{$\tau_T$ measurement} Normalized $I_T$ transmission intensity curves after the voltage reversal, in the O1 cell. $\tau_T$ is obtained as the time of minimum intensity (dots), for each voltage.} 
\end{figure}

\newpage
\subsection*{SI Section 5: Derivation of the kinetic behavior from the PCG model}

\label{app_kinetic_model}
Following the PC-Goldstone model from Clark et al. \cite{Diel_Clark2024}, we can discuss the re-orientation process of the polar fluid depending on the angle $\phi$ between $\Vec{P}$ and $\Vec{E}_{N_F}$. We can write the temporal evolution of $\phi$ as follows:
\begin{equation}
\begin{split} \label{eq_kin_generale}
    \frac{d\phi}{dt}&=\frac{1}{\gamma}PE_{N_F}\cdot sin(\phi)=-\frac{PV_{N_F}}{\gamma d} sin(\phi)=\\
    &=-\frac{P}{\gamma d}(\Delta V -cos(\phi)  V_{sat})sin(\phi)=\\
    &=-\Bigl[\tau_{high}^{-1}-cos(\phi) \tau_{low}^{-1}\Bigr] sin(\phi),
\end{split}
\end{equation}

wherein we exploited that $\Delta V$ is constant during the reversal process, since we have a squared voltage wave, whereas in \cite{Diel_Clark2024} a triangular wave is considered. Two characteristic times are defined as $\tau_{low}=\frac{\gamma d}{P  V_{sat}}$ and
$\tau_{high}=\frac{\gamma d}{P \Delta V}$, depending on the rotational viscosity $\gamma$.

\bigbreak

\paragraph{a) Derivation at low voltages}

\hfill \break

For $\Delta V \ll  V_{sat}$, in the Clark et al. model \cite{Diel_Clark2024}, the reversal is considered at angles $\phi\approx \pi/2$. Solving the problem for $\psi=\phi-\pi/2\approx 0$, and thus $sin(\psi)\approx \psi$ and $sin(\psi)\approx 1$, the Eq. \ref{eq_kin_generale} can be written as follows:
\begin{equation}
    \dot {\psi}\approx-\tau_{high}^{-1}-\psi         \cdot \tau_{low}^{-1},
\end{equation}
which is a first-order linear ordinary differential equation. The solution is the following:

\begin{equation}
    \psi(t)=ke^{-t/\tau_{low}}-\frac{\tau_{low}}{\tau_{high}}=ke^{-t/\tau_{low}}-\frac{\Delta V}{ V_{sat}},
\end{equation}

meaning that we have an exponential relaxation, with characteristic time $\tau_{low}$, to an equilibrium condition $\psi(t)=-\frac{\Delta V}{ V_{sat}}$. $k$ is determined by the initial condition of the problem $\psi(t=0)$. Also the polarization $P_z$ faced to the electrode as the same exponential behavior:

\begin{equation}
    \frac{P_z(t)}{P}=cos(\phi(t))=-sin(\psi(t))\approx-ke^{-t/\tau_{low}}+\frac{\Delta V}{ V_{sat}},
\end{equation}

whose equilibration as $P_z(t)/P=\Delta V/ V_{sat}$ satisfies the $E_{N_F}=0$ condition for $\Delta V< V_{sat}$.

We can also calculate the current intensity in the polarization reversal process as follows:

\begin{equation}
    \frac{I(t)}{PS}=\frac{1}{P}\frac{dP_Z}{dt}=+\frac{k}{\tau_{low}}e^{-t/\tau_{low}},
\end{equation}

that is again an exponential decay with with characteristic time $\tau_{low}$. This quantity can be referred to the resistivity as $\tau_{low}=\frac{\gamma d}{P V_{sat}}=\frac{\rho_{N_F}\cdot P d}{ V_{sat}}=R_{eff}\cdot C_{eff}$,  which is the same found in our model for the PPR wall motion.
\bigbreak

\paragraph{b) Derivation at high voltages} 

\hfill \break

For $\Delta V \gg  V_{sat}$, i.e., $\tau_{high}^{-1} \gg \tau_{low}^{-1}$, the Eq. \ref{eq_kin_generale} can be written as follows:
\begin{equation}
    \dot {\phi}\approx-\tau_{high}^{-1}sin(\phi),
\end{equation}
which is a first-order nonlinear ordinary differential equation, whose solution is the following:
\begin{equation}
    \phi(t)=2 cot^{-1}(e^{k+t/\tau_{high}}),
\end{equation}
wherein $cot^{-1}(x)$ is the inverse of the cotangent function and $k$ is determined by the initial condition $\phi(t=0)$. Starting from $\phi(t=0)\approx \pi$, $\phi(t)$ exhibits a smooth transition to $\phi\approx 0$ for long times, similarly to a sigmoidal trend. The resulting $P_z(t)=Pcos(\phi(t))$ goes from $-P$ to $P$, and we can find the expression of the current intensity:
\begin{equation}
    \frac{I(t)}{PS}=\frac{1}{P}\frac{dP_Z}{dt}=\frac{2 e^{k+x/\tau_{high}}\cdot sin[2 cot^{-1}(e^{k+t/\tau_{high}})] }{\tau_{high}[e^{2(k+t/\tau_{high})}+1]},
\end{equation}
that is a symmetric bell shape with peak at $t=k\cdot \tau_{high}=k \frac{\rho_{N_F}\cdot P d}{\Delta V}$, which scales linearly with $\Delta V^{-1}$.


\bibliography{references}

\bibliographystyle{unsrt}

\end{document}